\documentclass[submission, Phys]{SciPost}

\hypersetup{
    colorlinks,
    linkcolor={red!50!black},
    citecolor={blue!50!black},
    urlcolor={blue!80!black}
}

\usepackage[bitstream-charter]{mathdesign}
\usepackage{slashed}
\DeclareSymbolFont{usualmathcal}{OMS}{cmsy}{m}{n}
\DeclareSymbolFontAlphabet{\mathcal}{usualmathcal}

\begin{document}

\begin{center}{\Large \textbf{ Electromagnetic corrections in hadronic tau decays
\\
}}\end{center}

\begin{center}
Alejandro Miranda\textsuperscript{1$\star$} \end{center}

\begin{center}
{\bf 1} Institut de Física d’Altes Energies (IFAE) and
The Barcelona Institute of Science and Technology,
Campus UAB, 08193 Bellaterra (Barcelona), Spain.
\\[\baselineskip]
$\star$ \href{mailto:jmiranda@ifae.es}{\small jmiranda@ifae.es}
\end{center}

\begin{center}
\date{}
\end{center}


\definecolor{palegray}{gray}{0.95}
\begin{center}
\colorbox{palegray}{
  \begin{tabular}{rr}
  \begin{minipage}{0.1\textwidth}
    \includegraphics[width=30mm]{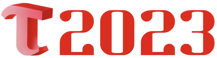}
  \end{minipage}
  &
  \begin{minipage}{0.81\textwidth}
    \begin{center}
    {\it The 17th International Workshop on Tau Lepton Physics}\\
    {\it Louisville, USA, 4-8 December 2023} \\
    \doi{10.21468/SciPostPhysProc.?}\\
    \end{center}
  \end{minipage}
\end{tabular}
}
\end{center}

\section*{Abstract}
{\boldmath\textbf{We briefly review electromagnetic radiative corrections in semileptonic tau decays and their main applications.}
}

\vspace{\baselineskip}



\vspace{10pt}
\noindent\rule{\textwidth}{1pt}
\tableofcontents
\noindent\rule{\textwidth}{1pt}
\vspace{10pt}


\section{Introduction}
\label{sec:intro}
The tau lepton is the only one massive enough to decay into hadrons, making it a valuable tool for studying the hadronization of QCD at low energies in rather clean conditions \cite{Pich:2013lsa}. Table \ref{Tab:1} summarizes the branching fraction precision of the main hadron tau decay channels, the knowledge of the corresponding radiative corrections (RadCors) and the main applications of these analyses. We used: LFU (Lepton Flavor Universality) and NSI (non-standard interactions), $V_\text{us}$ enters Cabibbo unitarity tests.

\begin{table}[h!]
    \centering
    \begin{tabular}{|c|c|c|c|}
    \hline
        $H^-$ & Branching ratio precision \cite{ParticleDataGroup:2022pth} & RadCors & Application(s)\\
        \hline
 $\pi^-$ & $0.5\%$ & \cite{Decker:1994dd,Decker:1994ea,Arroyo-Urena:2021nil,Arroyo-Urena:2021dfe} & LFU, NSI\\
  $K^-$ & $1.4\%$ & \cite{Decker:1994dd,Decker:1994ea,Arroyo-Urena:2021nil,Arroyo-Urena:2021dfe} & $V_{us}$, LFU, NSI\\
  $\pi^-\pi^0$ & $0.4\%$ & \cite{Cirigliano:2001er,Cirigliano:2002pv,Flores-Tlalpa:2005msx,Flores-Baez:2006yiq,Miranda:2020wdg,GutierrezSantiago:2020bhy,Masjuan:2023qsp,Escribano:2023seb} & $\rho^{(\prime)},\,(g-2)_\mu$, NSI \\ $K^-K^0$ & $2.3\%$ & \cite{Escribano:2023seb} & $\rho^\prime$, NSI\\
 $\bar{K}^0\pi^-$ & $1.7\%$ & \cite{Antonelli:2013usa,Flores-Baez:2013eba,Escribano:2023seb} & $K^*$, $V_{us}$, CPV, NSI\\
 $K^-\pi^0$ & $3.5\%$ & \cite{Antonelli:2013usa,Flores-Baez:2013eba,Escribano:2023seb} & $K^*$, $V_{us}$,  NSI\\
 $K^-\eta$ & $5.2\%$ & \cite{Escribano:2023seb} & $K^{*(\prime)}$, NSI \\
 $\pi^-\pi^+\pi^-$ & $0.5\%$ & x & $a_1$, NSI\\
 $\pi^-\pi^-\pi^0$ & $1.1\%$ & x & $a_1$, NSI\\
         \hline
    \end{tabular}
    \caption{Main semileptonic tau decay channels, precision of their measurement, RadCors available (x when missing) and main applications. Short-distance corrections were computed in Refs.~\cite{Sirlin:1977sv, Marciano:1988vm}.}
    \label{Tab:1}
\end{table}
The electromagnetic RadCors require the inclusion of virtual and real photons. The structure-independent (SI) contributions to the $K\pi$ channel were studied in Refs.~\cite{Antonelli:2013usa,Flores-Baez:2013eba}. In Ref.~\cite{Escribano:2023seb} we first computed the structure-dependent (SD) corrections for these decays and the remaining two-meson modes. Virtual photon corrections are IR divergent and induce a shift (and a dependence on an additional variable, $u$, due to the four-body kinematics) to the form factors, which was studied -within Chiral Perturbation Theory, $\chi PT$ \cite{Weinberg:1978kz,Gasser:1983yg,Gasser:1984gg}- in Ref.~\cite{Cirigliano:2001er}. We recall that the SI part of the radiative process is introduced via Low's theorem \cite{Low:1954kd}, so that the leading term in the photon low-energy expansion is fully determined by the non-radiative decay amplitude.

\section{Amplitude, observables, RadCors and new physics tests}
The most general amplitude for the processes $\tau^-(P)\to P^-(p_-) P^0(p_0) \nu_\tau(q) \gamma(k)$ is given by \cite{Guevara:2016trs, Miranda:2020wdg}
\begin{equation}
\mathcal{M}=\frac{eG_\text{F}V_{ud}^*}{\sqrt{2}}\epsilon_\mu^*\Bigg\lbrace\frac{H_\nu(p_-,p_0)}{k^2-2k\cdot P}\bar{u}(q)\gamma^\nu(1-\gamma_5)(M_\tau+\slashed{P}-\slashed{k})\gamma^\mu u(P)+(V^{\mu\nu}-A^{\mu\nu})\bar{u}(q)\gamma_\nu(1-\gamma_5)u(P)\Bigg\rbrace,
\end{equation}
where the hadron matrix element is
\begin{equation}
H^\nu(p_-,p_0)=C_V F_+(t)Q^\nu+C_S\frac{\Delta_{-0}}{t}q^\nu F_0(t)\,,\quad t=q^2\,,
\end{equation}
with $q^\nu=(p_-+p_0)^\nu$, $Q^\nu=(p_--p_0)^\nu-\frac{\Delta_{-0}}{t}q^\nu$ and $\Delta_{ij}=m_i^2-m_j^2$. The vector and axial-vector contributions can be split into the SI and SD parts, according to the Low and Burnett-Kroll \cite{Low:1954kd,Burnett:1967km} theorems: $V^{\mu\nu}=V^{\mu\nu}_\text{SI}+V^{\mu\nu}_\text{SD},\;A^{\mu\nu}=A^{\mu\nu}_\text{SD}$ at leading order in $\chi PT$ and fulfill $k_\mu A^{\mu\nu}=0$, $k_\mu V^{\mu\nu}=H^\nu(p_-,p_0)$ due to gauge invariance. For the $K^-\pi^0$ channel $C_V=C_S=1/\sqrt{2}$, the coefficients for the other modes can be checked in Ref.~\cite{Escribano:2023seb}. The required form factors that we use have been constructed, within a dispersive framework, in Refs.~\cite{GomezDumm:2013sib, Escribano:2013bca, Escribano:2014joa, Escribano:2016ntp,Gonzalez-Solis:2019iod}.
At leading order, the $V^{\mu\nu}$ are saturated by the exchange of (axial-)vector resonances. QCD short-distance constraints specify the resonance couplings that contribute up to next-to-leading order in $\chi PT$ in terms of the pion decay constant, $F$ \cite{Ecker:1988te, Ecker:1989yg}: $F_V=\sqrt{2}F,\,G_V=F/\sqrt{2},F_A=F$. In order to estimate the uncertainty due to missing higher chiral orders we also consider the relations that would be obtained adding the couplings at the next order \cite{Cirigliano:2006hb, Kampf:2011ty, Roig:2013baa} (which allow to comply with short-distance QCD not only for $2-$ but also for $3-$point Green functions and related form factors), that include $F_V=\sqrt{3}F,\,G_V=F/\sqrt{3},\,F_A=\sqrt{2}F$, and take the difference between the results with either set of constraints as a measure of our model-dependent uncertainty. To isolate the SD effects in the decay rates, we consider $E_\gamma^{\mathrm{cut}}\geq 300$ MeV in the decay spectra shown in Ref.~\cite{Escribano:2023seb} (in the $K^-K^0$ mode they are important even below $100$ MeV, given the kaon masses). In any case, the Low's approximation is insufficient to describe these decays for $E_\gamma\geq 100$ MeV where SD effects dominate. We also predict that the relation between the $\bar{K}^0\pi^-/K^-\pi^0$ branching fractions $\sim 2m_K/m_\pi$ in the Low limit, gets substantially modified for larger photon energies due to an accidental cancellation among contributions from different orders in $E_\gamma$ in the $\bar{K}^0\pi^-$ case. Measuring the photon/di-meson spectrum in this channel (cutting the lowest energy photons) would then be an important feedback for our SD input.\\

The photon-inclusive double differential decay rate can be written as
\begin{equation}\begin{split}\label{eq.spectrum}
\frac{\mathrm{d}\Gamma}{\mathrm{d}t}\Big|_{P^-P^0(\gamma)}=&\frac{G_\text{F}^2|V_{uD}F_+(0)|^2S_\text{EW}M_\tau^3}{768\pi^3t^3}\left(1-\frac{t}{M_\tau^2}\right)^2\lambda^{1/2}(t,m_-^2,m_0^2)G_\text{EM}(t)\\[1ex]
&\times\Big[C_V^2|\tilde{F}_+(t)|^2\left(1+\frac{2t}{M_\tau^2}\right)\lambda(t,m_-^2,m_0^2) + 3C_S^2\Delta_{-0}^2|\tilde{F}_0(t)|^2\Big]\,,
\end{split}\end{equation}
where the function $G_\text{EM}(t)$ \cite{Cirigliano:2001er} encodes the long-distance electromagnetic RadCors, $D=d,s$ and the tilded form factors are normalized to $F_+(0)$.
For simplicity, we split the contributions to the decay width as
\begin{equation}
\frac{\mathrm{d}\Gamma}{\mathrm{d}t}\Big|_{P^-P^0(\gamma)}=\frac{\mathrm{d}\Gamma}{\mathrm{d}t}\Big|_{P^-P^0}+\frac{\mathrm{d}\Gamma}{\mathrm{d}t}\Big|_{III}+\frac{\mathrm{d}\Gamma}{\mathrm{d}t}\Big|_{IV/III}+\frac{\mathrm{d}\Gamma}{\mathrm{d}t}\Big|_{\mathrm{rest}}\,,
\end{equation}
where the first two terms define the $G_\text{EM}^{(0)}(t)$ (namely the leading Low approximation plus non-radiative contributions), the third one is negligible and the last one gives the remainder of the $G_\text{EM}(t)$, which we call $\delta G_\text{EM}(t)$. The second and third term correspond to the Low approximation, where we separated $(IV/III)$ the phase space accessible through the four-body decay but not through the three-body one.\\

There are two models in the literature for incorporating the RadCors into the form factors. Differences between them are negligible in kaon decays, where they were introduced \cite{Cirigliano:2001mk,Cirigliano:2008wn}, but we have found this no longer holds in tau decays \cite{Escribano:2023seb}. Indeed, the dominant source of uncertainty of our RadCors comes from the difference between both factorization models in defining $F_{+/0}(t,u)=F_{+/0}(t)+\delta F_{+/0}(t,u)$, where $\delta F_0(t,u)=\delta F_+(t,u)+\frac{t}{\Delta_{-0}}\delta\bar{f}_-(u)$. We just quote here our preferred factorization approach~\footnote{Reasons are discussed in detail in Ref.~\cite{Escribano:2023seb}. Essentially, it warrants smoother RadCors, which is physically expected, by construction.}, according to which
\begin{equation}
\frac{\delta F_+(t,u)}{F_+(t)}=\frac{\alpha}{4\pi}\left[2(m_-^2+M_\tau^2-u)\mathcal{C}(u,M_\gamma)+2\log\left(\frac{m_-M_\tau}{M_\gamma^2}\right)\right]+\delta\bar{f}_+(u)\,,
\end{equation}
including the regulator for the photon mass, which cancels in all observables, permitting to take the vanishing $M_\gamma$ limit straightforwardly.\\

Integrating Eq.~(\ref{eq.spectrum}) upon $t$ gives the partial decay width
\begin{equation}
\Gamma_{P^-P^0(\gamma)}=\frac{G_\text{F}^2S_\text{EW}M_\tau^5}{96\pi^3}\left\vert V_{uD}F_+(0)\right\vert^2 I^\tau_{P^-P^0}(1+\delta^{P^-P^0}_\text{EM})^2\,,
\end{equation}
which defines the RadCor $\delta^{P^-P^0}_\text{EM}$, with
\begin{equation}\begin{split}
I^\tau_{P^-P^0}=&\frac{1}{8M_\tau^2}\int_{t_{thr}}^{M_\tau^2}\frac{\mathrm{d}t}{t^3}\left(1-\frac{t}{M_\tau^2}\right)^2\lambda^{1/2}(t,m_-^2,m_0^2)\Bigg[C_V^2|\tilde{F}_+(t)|^2\left(1+\frac{2t}{M_\tau^2}\right)\lambda(t,m_-^2,m_0^2)\\[1ex]
&+3C_S^2\Delta_{-0}^2|\tilde{F}_0(t)|^2\Bigg]\,.
\end{split}\end{equation}
\\
Our main results correspond to the following RadCors (expressed always in $\%$)
\begin{eqnarray}\label{eq.mainresults}
&&\delta_\text{EM}^{K^-\pi^0}=-\left(0.009^{+0.010}_{-0.118}\right)\,,\quad \delta_\text{EM}^{\bar{K}^0\pi^-}=-\left(0.166^{+0.100}_{-0.157}\right)\,,\nonumber\\[1ex]
&&\delta_\text{EM}^{K^-K^0}=-\left(0.030^{+0.032}_{-0.180}\right)\,,\quad \delta_\text{EM}^{\pi^0\pi^-}=-\left(0.186^{+0.114}_{-0.203}\right)\,,
\end{eqnarray}
where the uncertainty due to the so far missing virtual photon SD corrections is taken into account (estimating it from the corresponding results in the one-meson modes, \cite{Arroyo-Urena:2021nil}). This correction will be presented elsewhere. As expected, from the mass-dependence on the radiating particle in the Low limit, the RadCors in Eqs. (\ref{eq.mainresults}) are considerably larger for the modes with a $\pi^-$ than for those with a $K^-$. In the latter, the uncertainty is completely asymmetric as it is dominated by the missing virtual SD correction (of known sign); while in the former, the asymmetry of the error is reduced since the uncertainty associated to the resonance couplings (with an effect of unknown sign) is non-negligible.
For completeness we also quote our estimates for the RadCors in the $K^-\eta^{(\prime)}$ modes, which were obtained exploiting the dominance of the vector ($\eta$) and scalar ($\eta^\prime$) form factor, respectively:
\begin{equation}\label{eq.KetaRadCors}
\delta_\text{EM}^{K^-\eta}=-\left(0.026^{+0.029}_{-0.163}\right)\,,\quad \delta_\text{EM}^{K^-\eta^\prime}=-\left(0.304^{+0.422}_{-0.185}\right)\,.
\end{equation}
The largest RadCor is obtained for the $K^-\eta^\prime$ mode due to the dominance of the scalar form factor in the decay spectra~\cite{Escribano:2013bca}, as the corresponding kinematical dependence enhances the effect of the RadCors.
Our results for the two-meson RadCors agree with earlier estimations (with improved precision) where available, and fill the gap for those yet uncomputed.\\

Together with our improved computation of the RadCors in the one-meson tau decays \cite{Guo:2010dv,Guevara:2013wwa,Guevara:2021tpy,Arroyo-Urena:2021nil,Arroyo-Urena:2021dfe}, the results presented here enable more precise new physics tests using a low-energy Effective Field Theory of the $\tau^-\to\bar{u}D\nu_\tau$ decays \cite{Cirigliano:2009wk}, which has been exploited in Refs.~\cite{Garces:2017jpz, Cirigliano:2017tqn,Miranda:2018cpf,Cirigliano:2018dyk,Rendon:2019awg,Chen:2019vbr,Gonzalez-Solis:2019lze,Gonzalez-Solis:2020jlh,Chen:2020uxi,Chen:2021udz,Cirigliano:2021yto,Arteaga:2022xxy} (see Ref.~\cite{RoigGarces:2023vly} for a more detailed summary than the one presented here). Particularly, in Ref.~\cite{Escribano:2023seb} we update our fits for either $\Delta S=0$ and $\Delta S=1$ one- and two-meson tau decays as well as our joint fit assuming minimal flavor violation \cite{DAmbrosio:2002vsn} to allow for their combination, breaking thereby a degeneracy in the new physics parameter space. Interestingly, the inclusion of our RadCors increases the compatibility with the SM in the largest Wilson coefficient appearing in the strangeness-conserving channels (the one accounting for scalar non-standard interactions, $\epsilon_S^\tau$). The changes induced by our RadCors are, in all other instances, much smaller and covered by the uncertainties. Our dominant errors are statistical in the $|\Delta S|=1$ processes (we emphasize once again the importance of improving the measurements of the strange tau spectral function and their contributing channels), and theoretical both in the $\Delta S=0$ channels and in the joint analysis. Under the weak coupling assumption, our limits push the new physics affecting the $\tau^-\to\bar{u}D\nu_\tau$ processes to energies larger than a few TeVs. It will also be interesting to include our improved RadCors in updated studies of lepton universality and CKM unitarity using two-meson tau decays.


\section{Conclusions}
RadCors are needed to improve the precision of new physics analyses beyond the percent level. Here we have reviewed our evaluation of those entering two-meson tau decays and their application in searches for non-standard interactions. This information was available for the $\pi^-\pi^0$ case and only the SI part, for the $K\pi$ channels, was known before. We have filled this gap and computed the SD part stemming from real photons (the corresponding virtual contributions were only estimated and their calculation is in progress) in this work, which has also included the $K^-K^0$ and $K^-\eta^{(\prime)}$ modes. Our main results are the numerical values (in $\%$) of the RadCors in Eqs.~(\ref{eq.mainresults}) (see also Eq.~(\ref{eq.KetaRadCors})). These are in agreement with earlier publications and reduce the uncertainty band to a half, approximately. We have also put forward that the factorization prescription is very important in semileptonic tau  (contrary to the kaon case) decays and explained our preferred choice for it, accounting for the corresponding uncertainty in our results. The measurement of the spectra (for $E_\gamma\gtrsim100$ MeV, to be sensitive to the SD contributions) would help us to reduce our theory uncertainty substantially, benefiting all related new physics searches.

\section*{Acknowledgements}
The author acknowledges a pleasant collaboration with Rafel Escribano and Pablo Roig in the topic here discussed. Congratulations to Swagato and his team for the excellent organization of this workshop. 

\paragraph{Funding information}
A. M. is supported by MICINN with funding from European Union NextGenerationEU (PRTR-C17.I1). IFAE is partially funded by the CERCA program of the Generalitat de
Catalunya. This work has been partly financed by the Ministerio de Ciencia
e Innovación under grant PID2020-112965GB-I00, and by the Departament de Recerca i Universitats from Generalitat de Catalunya to the Grup de Recerca ‘Grup de Física Teòrica UAB/IFAE’
(Codi: 2021 SGR 00649).






\nolinenumbers

\end{document}